\documentclass{elsart}

\usepackage{natbib}   
\usepackage{amssymb}  
\usepackage{epsfig}

\def\url#1{{\ttfamily\def\/{/\discretionary{}{}{}}#1}}
\def\bibcode#1{(\texttt{#1})}

\newcommand{\Fion}{F_{\rm ion}}
\newcommand{\tH}{t_{\rm H}}
\newcommand{\Om}{\Omega_{M,0}}
\newcommand{\Ol}{\Omega_{\Lambda,0}}
\newcommand{\Ob}{\Omega_b}
\def\avg#1{\left\langle#1\right\rangle}
\def\rp#1{\left(#1\right)}

\newcommand{\Lam}{\Lambda}
\newcommand{\del}{\delta}

\newcommand{\hinv}{{h^{-1}}}

\newcommand{\himpc}{\hinv{\rm\,Mpc}}

\newcommand{\Msun}{M_{\odot}}
\newcommand{\himsun}{\hinv{\Msun}}

\def\Fig#1{Figure~\ref{#1}}

\begin{document}

\begin{frontmatter}

\title{Future Evolution of the Intergalactic Medium 
in a Universe Dominated by a Cosmological Constant}

\author{Kentaro Nagamine\thanksref{ken}},
\author{Abraham Loeb\thanksref{avi}}
\address{Harvard-Smithsonian Center for Astrophysics, 
60 Garden Street, MS 51, Cambridge, MA 02138}

\thanks[ken]{E-mail: knagamin@cfa.harvard.edu}
\thanks[avi]{E-mail: aloeb@cfa.harvard.edu}


\begin{abstract}
We simulate the evolution of the intergalactic medium (IGM) in a universe
dominated by a cosmological constant. We find that within a few Hubble
times from the present epoch, the baryons will have two primary phases: one
phase composed of low-density, low-temperature, diffuse, ionized gas which
cools rapidly with cosmic time due to adiabatic exponential expansion, and 
a second phase of high-density, high-temperature gas in virialized dark
matter halos which cools much more slowly by atomic processes.  The mass
fraction of gas in halos converges to $\sim 40$\% at late times, about
twice its calculated value at the present epoch.  We find that in a few
Hubble times, the large scale filaments in the present-day IGM will rarefy
and fade away into the low-temperature IGM, and only islands of virialized
gas will maintain their physical structure.  We do not find evidence for
fragmentation of the diffuse IGM at later times.  More than 99\% of the gas
mass will maintain a steady ionization fraction above 80\% within a few
Hubble times. The diffuse IGM will get extremely cold and dilute but remain
highly ionized, as its recombination time will dramatically exceed the age
of the universe.
\end{abstract}


\end{frontmatter}


\section{Introduction}
\label{section:intro}

Independent data sets, involving the temperature anisotropies of the cosmic
microwave background \citep{Berna00, Hanany00, WMAP}, the luminosity
distances of Type Ia supernovae \citep{Garnavich98, Riess98, Perlmutter98,
Hanany00}, the large-scale distribution of galaxies \citep{Peacock, Verde},
and cluster abundances \citep{Eke, Bahcall03}, appear to be all consistent
with a single set of cosmological parameters. In the concordance
$\Lam$--Cold Dark Matter ($\Lam$CDM) model, the universe is flat and its
expansion rate is currently accelerating; the cosmic mass density is
dominated ($\sim 70\%$) by the vacuum (the so-called cosmological constant
or ``dark energy'') with the remaining density mostly in the form of cold
dark matter ($\sim 26\%$) and baryons ($\sim 4\%$).

Given the emergence of a standard model in cosmology with a specific set of
parameters, it is of much interest to follow the immediate consequence of
these parameters in terms of the near future evolution of the $\Lam$CDM
universe. While several recent studies considered the qualitative
implications of the accelerating universe by analytic means \citep{AL97,
Krauss, Chiueh, Gud02, Loeb02}, it is clear that further quantitative
insight can be gained only through direct numerical simulations. In
\citet[][hereafter Paper I]{Nag03}, we simulated the evolution of nearby
large-scale structure using a constrained realization of the Local Universe
with only dark matter particles.  We have found that structure will freeze
within two Hubble times from the present epoch, and that the dark matter
halo mass function will not evolve subsequently. \citet{Busha} further
studied the generic evolution of the density profile around dark matter
halos embedded in an accelerating universe.

In this paper we extend previous numerical work and study the evolution of
the baryonic component of the universe using a hydrodynamic cosmological
simulation. In Section~\ref{section:simulation} we describe the
simulation. Images of the simulated gas mass distribution and gas
temperature distribution are presented in Section~\ref{section:visual}.  
We describe the global evolution of gas temperature and overdensity in
Section~\ref{section:2d}, and provide a more quantitative analysis based 
on the distribution functions of these quantities in
Section~\ref{section:dist-overd} and \ref{section:dist-temp}.  
The ionization fraction of cosmic gas is analyzed in
Section~\ref{section:ionizefrac}.  Finally, we summarize our main
conclusions in Section~\ref{section:conclusion}.


\section{Simulation}
\label{section:simulation}

We carried out a hydrodynamic simulation with the concordance cosmological
parameters ($\Om$, $\Ol$, $\Ob$, $h$, $\sigma_8$, $n$)=(0.3, 0.7, 0.04,
0.7, 0.9, 1.0) from a redshift $z=99$ (i.e. a scale factor
$a=(1+z)^{-1}=0.01$) through the present time ($z=0$, $a=1$), and up to 6
Hubble times into the future ($a=166$; $\sim 84$ billion years from today).
Because the present Hubble time $\tH\equiv 1/H_0=14$ Gyr is very close to
the current age of the Universe $t_0=13.5$ Gyr in the adopted cosmology,
the epoch of $a=166$ roughly corresponds to $t=t_0+6\tH \approx 7\tH$ from
the Big Bang ($t=0$), where a subscript zero denotes present-day values.

We use the updated version of the parallel tree Smoothed Particle
Hydrodynamics (SPH) code
GADGET\footnote[1]{http://www.MPA-Garching.MPG.DE/gadget/}\citep{Springel01}.
The comoving box size of our simulation is $50\himpc$, and the particle
number is $64^3$ for the gas and $64^3$ for the dark matter, with
corresponding particle masses $m_{\rm DM}=3.4\times10^{10}\himsun$ and
$m_{\rm gas}=5.3\times10^9\himsun$ (so that a group of $\sim 50$ particles
has a total mass comparable to that of the Milky Way galaxy). We adopt the
`entropy-conserving' formulation of SPH as described by \citet{SH02}.  The
simulation includes a standard treatment of the radiative cooling and
heating of \citet{Katz} assuming that the gas is optically thin and in
ionization equilibrium (see Section~\ref{section:ionizefrac} for the
discussion on the validity of this assumption at late times). The abundance
of different ionic species, including H$^0$, He$^0$, H$^+$, He$^+$, and
He$^{++}$, is computed by solving the network of equilibrium equations
self-consistently with a uniform ultra-violet (UV) background radiation of
a modified \citet{Haardt} spectrum and with complete reionization at $z\sim
6$ \citep{Dave99, Becker}.  The UV background (or, equivalently, the
photoionization rate $\Gamma$ of the IGM) is linearly extrapolated beyond
the present epoch, after which it quickly approaches zero.  We have tested
the validity of this assumption by carrying out another simulation with a
maximum UV background field, where the value of the photoionization rate
was set equal to the $z=0$ value and remained constant afterwards.  Even
for this maximum UV background field, the future thermal evolution of the
IGM remained similar.  The exact behavior of the UV background field after
$z=0$ is unimportant because the photoionization time-scale is rapidly
increasing to values much longer than the Hubble time. This follows from
the fact that the cosmic volume element (by which the ionizing photons are
diluted) grows as $\propto a^{3}$ and $a$ grows exponentially with cosmic
time, i.e. $a\propto \exp\{\sqrt{\Ol}H_0 t\}$ at $t\gg t_0$.  The evolution
of the neutral hydrogen mass density in a similar simulation run was
discussed by \citet{NSH03a}.

Feedback from star formation and supernovae is formulated as in
\citet{SH03a, SH03b}; the underlying assumptions are also reviewed in
\citet{NSH03b}. However, the details of the adopted star formation model
should not be very important for the results presented here since the
heating of the gas on large scales is dominated by the gravitationally-induced
shocks. Although the numerical resolution of our simulation is not adequate
for following the detailed star formation history of the Universe
\citep[see][]{SH03b}, it does capture well the large-scale properties of
baryons in the IGM which are the focus of this work.  Future simulation
runs with a higher resolution and an improved model of star formation and
supernova feedback could address more complicate issues, such as the
distribution of metals in the IGM.

Our simulation was performed at the Center for Parallel Astrophysical
Computing\footnote[2]{http://cfa-www.harvard.edu/cpac/} at Harvard-Smithsonian
Center for Astrophysics.


\section{Images}
\label{section:visual}

\begin{figure}
\begin{center}
\epsfig{file=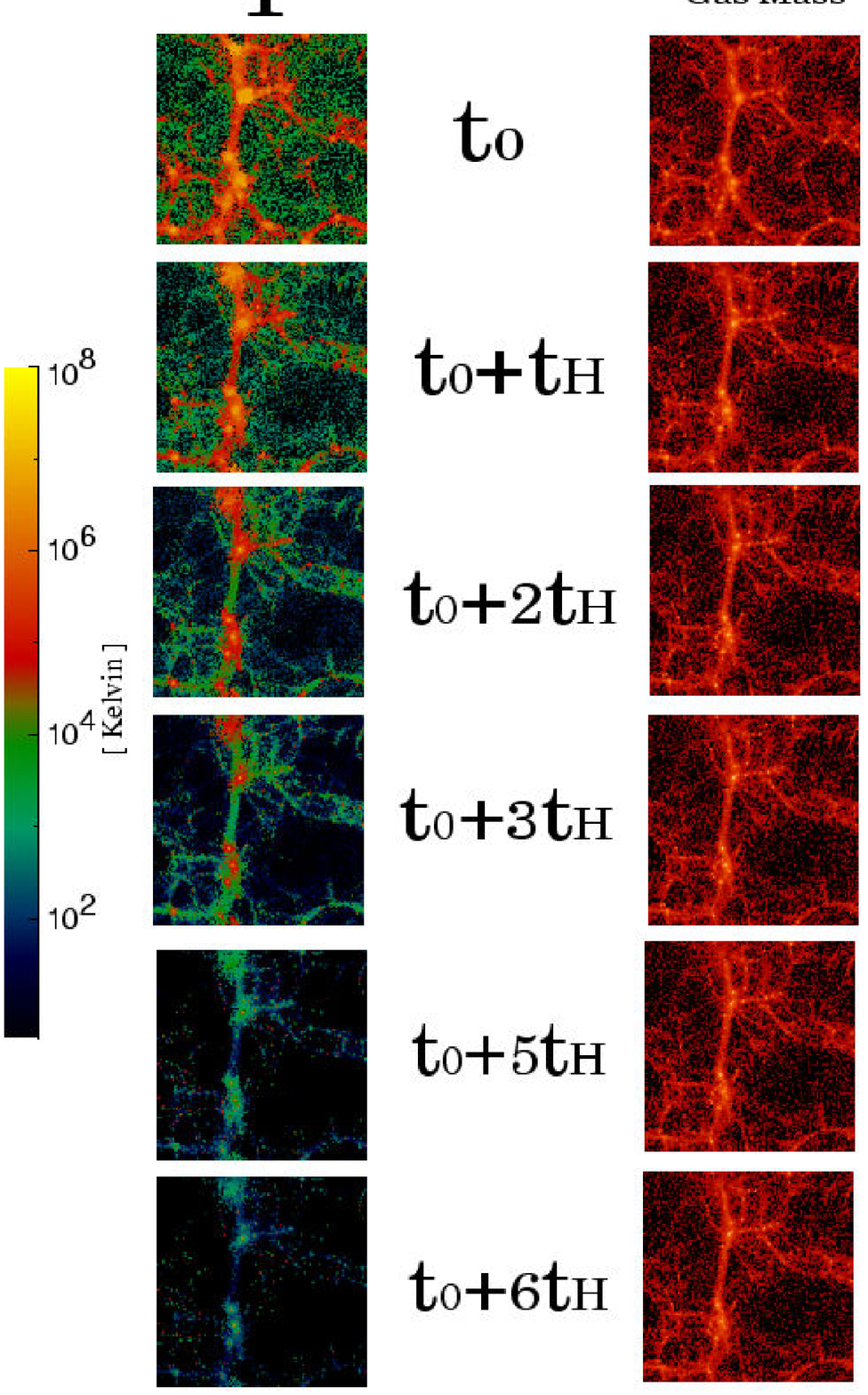,height=8in,width=5in, angle=0}
\caption{Projected distribution of mass-weighted gas temperature, $T$, in K
(left column) and projected surface gas mass density (right column) for a 
slice with a comoving thickness of $10\himpc$ and a comoving width of 
$50\himpc$ on a side.
\label{evolve.eps}}
\end{center}
\end{figure}

\Fig{evolve.eps} shows the projected distribution of mass-weighted
temperature (left column) and the mass surface density (right column) of
the gas as functions of cosmic time in a slice with a comoving thickness of
$10\himpc$ and a comoving width of $50\himpc$ on a side.

As shown in Paper I, the evolution of large-scale structure in the dark
matter distribution freezes within two Hubble times from the present
epoch. Similarly, the right column of \Fig{evolve.eps} indicates that the
gas distribution does not evolve either at late times.  The only noticeable
change in comoving coordinates relative to the present epoch is that the
filaments thin out as their mass is drained into X-ray clusters. The two
visible X-ray clusters (one at the top middle and the other at the bottom
middle of this simulation) are still connected by a low-temperature
filament of gas and dark matter even at $t=t_0+6\tH$.  In physical
coordinates, these filaments rarefy considerably as the universe expands
exponentially with cosmic time.

Due to the adiabatic expansion of the IGM, the temperature distribution of
the cosmic gas evolves dramatically with time.  At $t=t_0$, the
high-temperature clusters (with $T>10^7$~K, indicated by yellow) are
connected by filaments with a somewhat lower temperature ($10^5<T<10^7$~K,
indicated by red). As time progresses, the red filaments cool down
($T<10^5$~K in green), get thinner, and eventually disappear into the dark
background. Consequently, virialized dark matter halos with gas temperature
$T\ge 10^4$~K are left as `island universes' embedded within the
low-temperature IGM.

In principle, it is possible for the IGM to fragment into baryonic objects
(which do not necessarily overlap with dark matter halos) due to its rapid
adiabatic cooling.  We have tested for this possibility by applying the HOP
grouping algorithm \citep{HOP} to the gas particles in the simulation, but
found no evidence for such fragmentation.  We caution, however, that our
numerical resolution is not adequate for probing fragmentation below the
mass-scale of $M_{gas}\sim 5\times 10^{11}\himsun$.

\section{Growth Factor}

\begin{figure}
\begin{center}
\epsfig{file=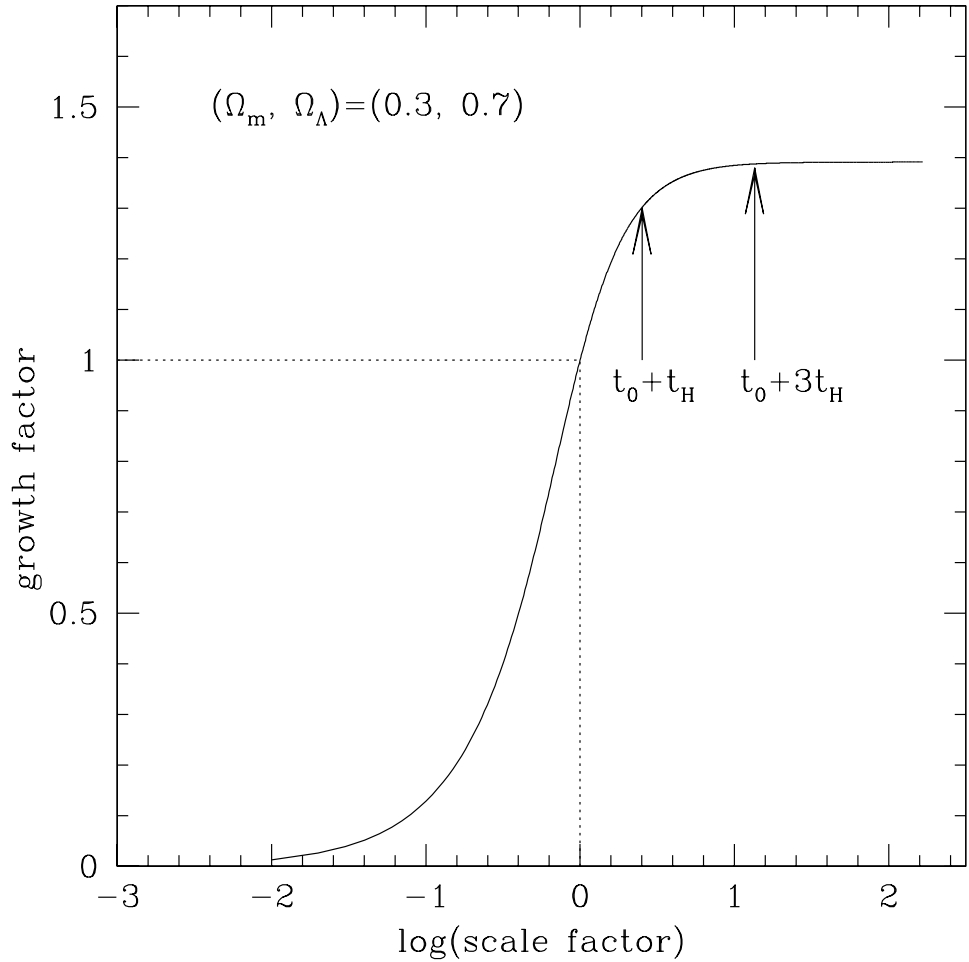,height=5in,width=5in, angle=0}
\caption{Growth factor as a function of the scale factor `$a$'.
The curve is normalized to unity at $a=1$, as indicated by the dotted line. 
The growth factor saturates within two Hubble times from the present epoch.
\label{growth.eps}}
\end{center}
\end{figure}

In order to demonstrate the freezing of the structure formation in the future
from a different perspective, we examine the evolution of the growth factor
of linear density fluctuations in a $\Lam$CDM universe.
\Fig{growth.eps} shows the growth factor \citep{Heath, Carroll} as a function 
of the logarithm of the scale factor, $a$, 
\begin{equation}
D(a) \propto \frac{1}{a}\frac{da}{d\tau}\int_0^a
\left(\frac{da'}{d\tau}\right)^{-3} da',
\end{equation}
where 
\begin{equation}
\left(\frac{da}{d\tau}\right)^2 = 1+\Om\left(\frac{1}{a}-1\right) +
\Ol(a^2-1),
\end{equation}
and $\tau = H_0 t$.
The curve is normalized to unity at $a=1$.  It clearly demonstrates that
the growth of structure ceases at about two Hubble times from the present
time in a Universe which is dominated by a cosmological constant,
consistently with the results of Paper I. This saturation is caused by the
exponential expansion of $a(t)$ as driven by the cosmological constant.


\section{Temperature -- Density Diagram}
\label{section:2d}

\begin{figure}
\begin{center}
\epsfig{file=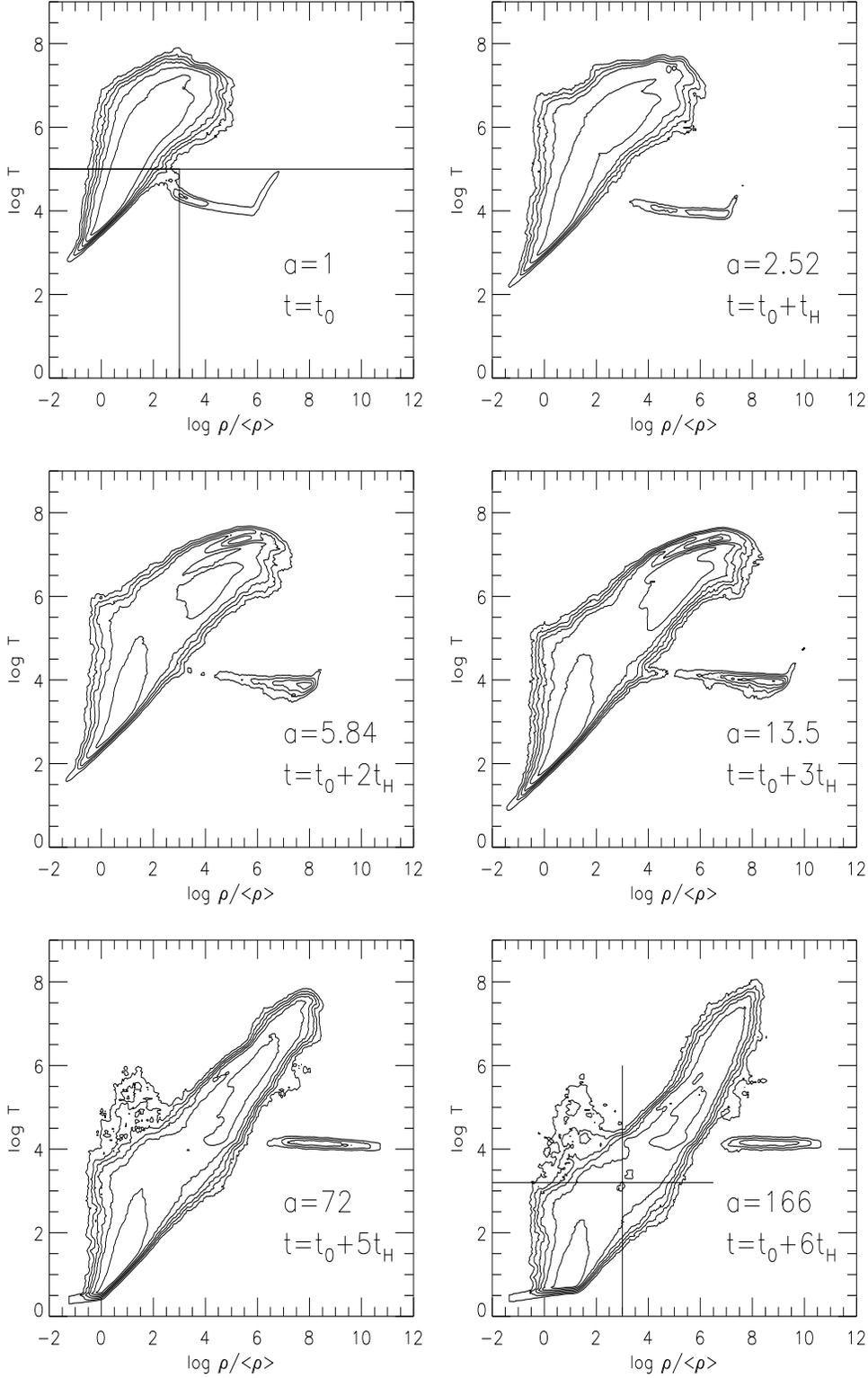,height=8in,width=5in, angle=0}
\caption{Snapshots of the mass-weighted gas temperature $T$ (in K) vs. gas 
density (relative to the mean value $\avg{\rho}$ at each epoch) for $t=t_0$, 
$t_0+\tH$, $t_0+2\tH$, $t_0+3\tH$, $t_0+5\tH$, and $t_0+6\tH$. The six 
contours are equally spaced in logarithmic intervals between the minimum and 
maximum values of the gas mass distribution on the plane.  See the text for 
the description of the solid lines in the top left and bottom right panels.
\label{trho.eps}}
\end{center}
\end{figure}

Next we project the global evolution of the gas properties on the
temperature--gas density ($T$--$\rho$) plane.  In \Fig{trho.eps} we show 
the mass-weighted distribution of baryons on the temperature --
gas overdensity plane as a function of cosmic time.  The time and
corresponding scale factor values are indicated in each panel, and the
plotted contours are equally spaced in six logarithmic intervals between
the minimum and maximum values on the plane. 

The top left panel shows the familiar geometry of the gas distribution at
the present time; the solid line divides the gas into 3 different
categories following the criteria of \citet{Dave01}. As shown later, the
dividing line of $\rho/\avg{\rho}\equiv (1+\delta) =10^3$ (where
$\avg{\rho}$ is the mean density) has a special meaning in terms of the
future evolution of the IGM as well.  The bottom left region of the diagram
corresponds to the diffuse IGM which has a low density and a low
temperature. Most of the gas mass in this region lies on the tight
nearly-adiabatic power-law relation, which can be described by analytic
means \citep{Hui}.  A plume-shaped region of shock-heated gas extends above
the horizontal solid line at $T=10^5$~K. The gas with temperatures
$10^5<T<10^7$~K is the so called {\it warm-hot intergalactic medium} (WHIM;
Cen \& Ostriker, 1999), and the dense gas with $T>10^8$~K is the {\it hot}
gas in clusters of galaxies. In the current study, the distinction between
the {\it warm-hot} and {\it hot} gas is not particularly important, and so
we simply refer to them both as {\it hot gas}.  Finally, the bottom right
portion of the diagram is occupied by condensed gas that has cooled
radiatively inside dark matter halos. 

Within two Hubble times into the future ($t=t_0+2\tH$), the innermost
contour line splits into two islands, one for the low-density diffuse IGM
and the other for the gas in virialized dark matter halos. 
The overdensity of gas in dense regions increases with time because the
physical density there remains nearly fixed while the background density
rarefies rapidly due to its Hubble expansion. 

The $T$--$\rho$ diagram is stretched into a more elongated shape in the 
distant future. This elongation of the contour can be understood by 
considering the gas in clusters of galaxies that have $T=10^{7.5}$~K and 
overdensity of $200$ at $z=0$. The Bremsstrahlung (which dominates the 
cooling rate at this temperature) cooling time of such gas is much longer 
than the current Hubble time:
\begin{equation}
t_{cool}\sim \frac{2\cdot \frac{3}{2}kT}{\epsilon_B} \sim 10^2
\rp{\frac{200}{\rho / \avg{\rho}_0}}\left({T\over 10^{7.5}~{\rm
K}}\right)^{1/2} \tH,
\label{eq:cooling}
\end{equation}
where $\epsilon_B\propto T^{0.5}\rho$ is the Bremsstrahlung cooling rate
per electron at a temperature $T$ and density $\rho$ of the cluster gas.
The factor of 2 in the numerator accounts for the thermal energy of the
protons in addition to that of the cooling electrons.
Because of this long cooling time, such gas in the outskirts of clusters 
of galaxies will remain hot in the distant future.

The contour peak at $T\sim 10^6$~K and $\log(\rho/\avg{\rho}) > 3.0$
corresponds to gas inside galaxies and groups of galaxies.  For this gas,
the significance of metal line cooling increases with time as the gas gets
enriched with supernova ejecta, and the cooling time of the enriched
gas is already shorter than the Hubble time at present.
Since our simulation does not include metal cooling
and star formation is not treated accurately due to lack of numerical
resolution, we cannot rule out the possibility that the peak at $T\sim
10^6$~K disappears at late times.

After six Hubble times ($t=t_0+6\tH$), the flat bottom of the contour 
plot reaches the temperature floor of $T=5$~K, which is set by hand in 
the simulation and is not physical.  The virialized gas maintains a
temperature above $T=10^4$~K because its cooling curve drops sharply
below this temperature. The vertical solid line indicates the dividing
overdensity of $10^3$. The horizontal solid line at $T=10^{3.2}$K will be
discussed later.


\section{Distribution \& Evolution of Gas Overdensity}
\label{section:dist-overd}

\begin{figure}
\begin{center}
\resizebox{9.5cm}{!}{\includegraphics{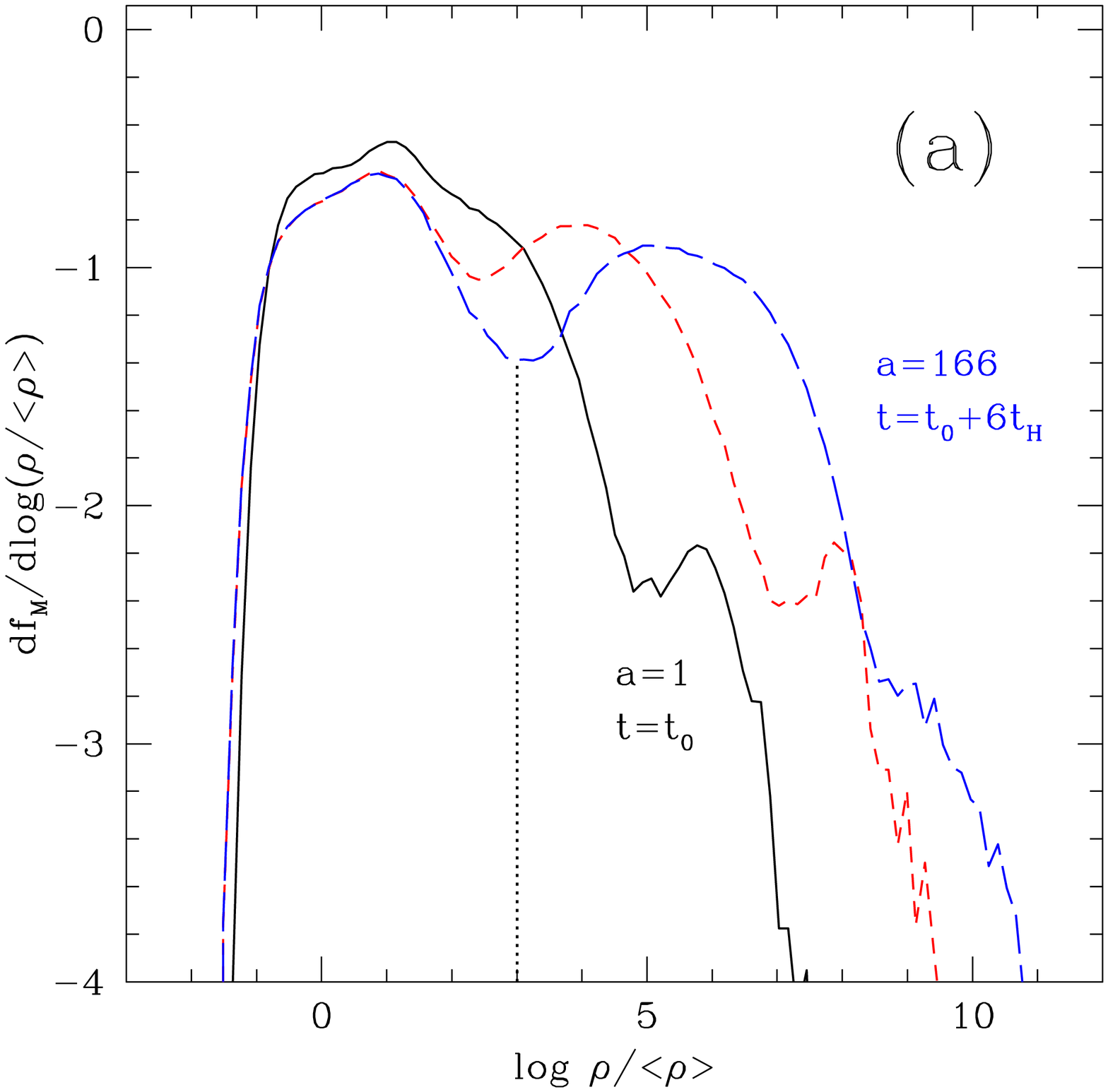}}\\
\hspace{0.3cm}
\resizebox{9.5cm}{!}{\includegraphics{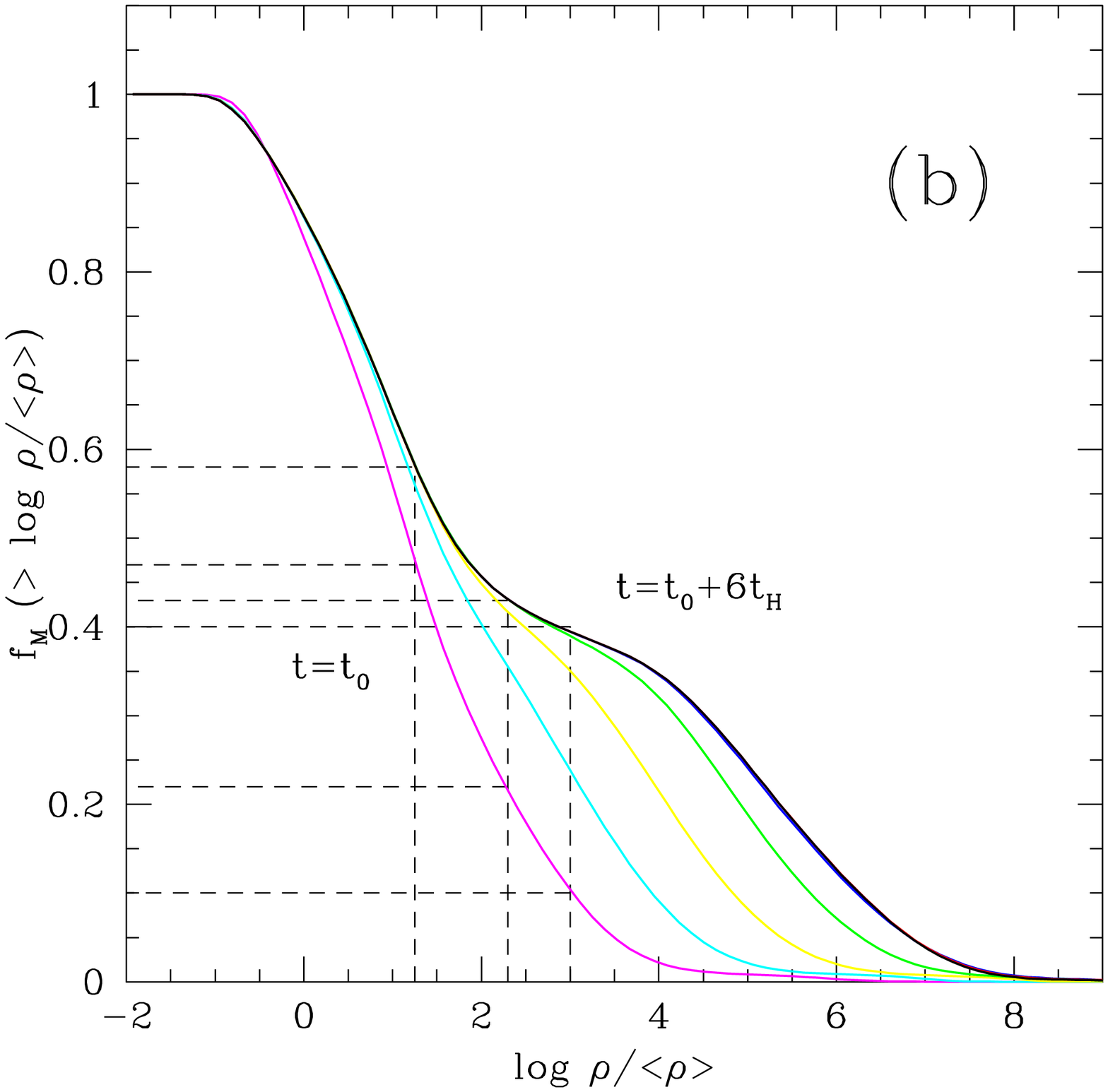}}\\%
\caption{{\it Panel (a):} Mass-weighted, differential distribution function
of gas overdensity (relative to the mean density $\avg{\rho}$ at each 
epoch), i.e. $df_M/d\log(\rho/\avg{\rho})$, for $t=t_0$, $t_0+2\tH$, and 
$t_0+6\tH$ from left to right.  
Hereafter we denote the gas mass fraction in general by $f_M$. 
A density contrast of $10^3$ is indicated by the dotted line which
splits the distribution into two regimes at $t=t_0+6\tH$. {\it Panel (b):}
Cumulative gas mass fraction with overdensity above a specific value, 
i.e. $f_M(> \rho/\avg{\rho})$, for $t=t_0$, $t_0+2\tH$, $t_0+3\tH$, $t_0+4\tH$, 
$t_0+5\tH$ and $t_0+6\tH$, from left to right. The characteristic values
to note are indicated by the dashed lines, and are described in the text.
\label{density.eps}}
\end{center}
\end{figure}

To describe the evolution of the IGM more quantitatively, we show in
\Fig{density.eps} the mass-weighted distribution function of the cosmic gas
as a function of its density contrast $df_M/d\log(\rho/\avg{\rho})$ 
[panel (a)], and the cumulative mass fraction of gas with overdensity larger 
than a specific value $f_M(> \rho/\avg{\rho})$ [panel (b)]. Hereafter, we 
denote the gas mass fraction in general by $f_M$. 

Panel (a) indicates that the gas distribution becomes bimodal as time
progresses; the bottom of the valley that separates the gas into two
islands is located near a density contrast $\rho/\langle\rho\rangle \equiv
1+\delta=10^3$ at $t=t_0+6\tH$, as indicated by the vertical dotted line.
This makes $1+\delta=10^3$ a natural choice for the borderline separating
the gas into its two phases: the diffuse IGM and the phase which remains
bound in virialized dark matter halos (as marked by the solid lines in
\Fig{trho.eps}).

\begin{figure}
\vspace{3cm}
\begin{center}
\epsfig{file=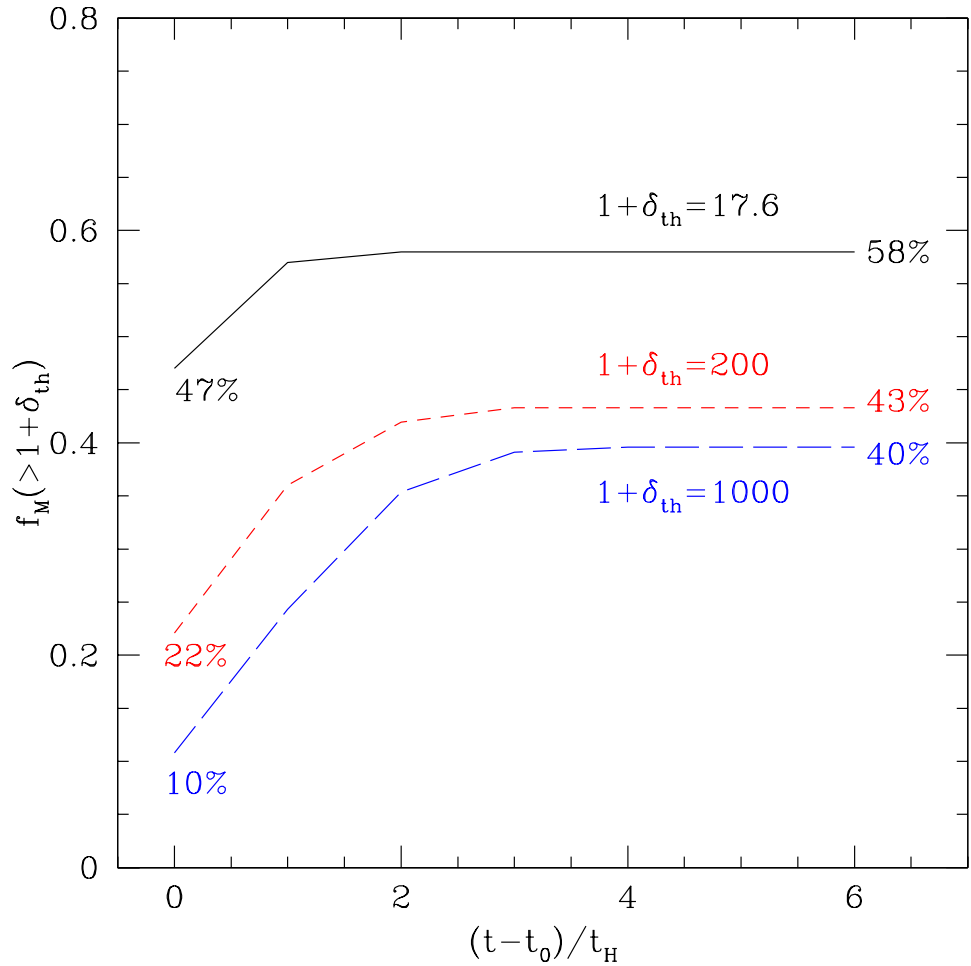,height=5in,width=5in, angle=0}
\caption{Gas mass fraction $f_M(>1+\delta_{\rm th})$ with density contrast 
higher than $1+\delta_{th}=\rho/\avg{\rho}=17.6$ (solid), 200 (short-dashed), 
and 1000 (long-dashed line), as a function of cosmic time from the present 
epoch in units of the current Hubble time.  The initial and final values of 
the gas mass fractions over the plotted range of time are mentioned next to 
each of the curves.
\label{boundfrac.eps}}
\end{center}
\end{figure}

Panel (b) shows that more gas shifts toward higher overdensities as it
falls into the potential wells of dark matter halos at late times. 
This is partially because the physical density in virialized halos remains 
nearly the same while the background density rarefies rapidly due to the 
exponential expansion of the universe. 
From this panel, we derive the gas mass fraction that is 
in the region above some overdensity value, as indicated by the dashed lines. 
We find that the gas mass fraction with $1+\delta > 10^3$ is 10\% (40\%) at 
$t=t_0$ ($t=t_0+6\tH$). The gas mass fraction with $1+\delta>200$ is 22\% 
(43\%) at $t=t_0$ ($t=t_0+6\tH$).  Finally, the fraction with $1+\delta>17.6$ 
is 47\% (58\%) at $t=t_0$ ($t=t_0+6\tH$).  These results are summarized in
\Fig{boundfrac.eps}, which shows the gas mass fraction 
$f_M(>1+\delta_{\rm th})$ with density contrast higher than 
$1+\delta_{th}=\rho/\avg{\rho}=17.6$ (solid), 200 (short-dashed), 
and 1000 (long-dashed line), as a function of cosmic time from the present 
epoch in units of the current Hubble time. This figure shows the convergence 
of the amount of gas that is trapped inside virialized dark matter halos in 
two Hubble times from the present time. The gas mass fractions with 
$1+\delta>200$ and $>1000$ both converge to $\sim 40\%$ at $t=t_0+6\tH$.  
At the lower threshold overdensity of $1+\delta=17.6$, the mass fraction 
increases to 58\%.  The threshold value of 17.6 was derived in Paper I as the 
critical overdensity above which mass remains bound to virialized objects at 
late times.


\section{Distribution \& Evolution of Gas Temperature}
\label{section:dist-temp}

\begin{figure}
\begin{center}
\resizebox{10cm}{!}{\includegraphics{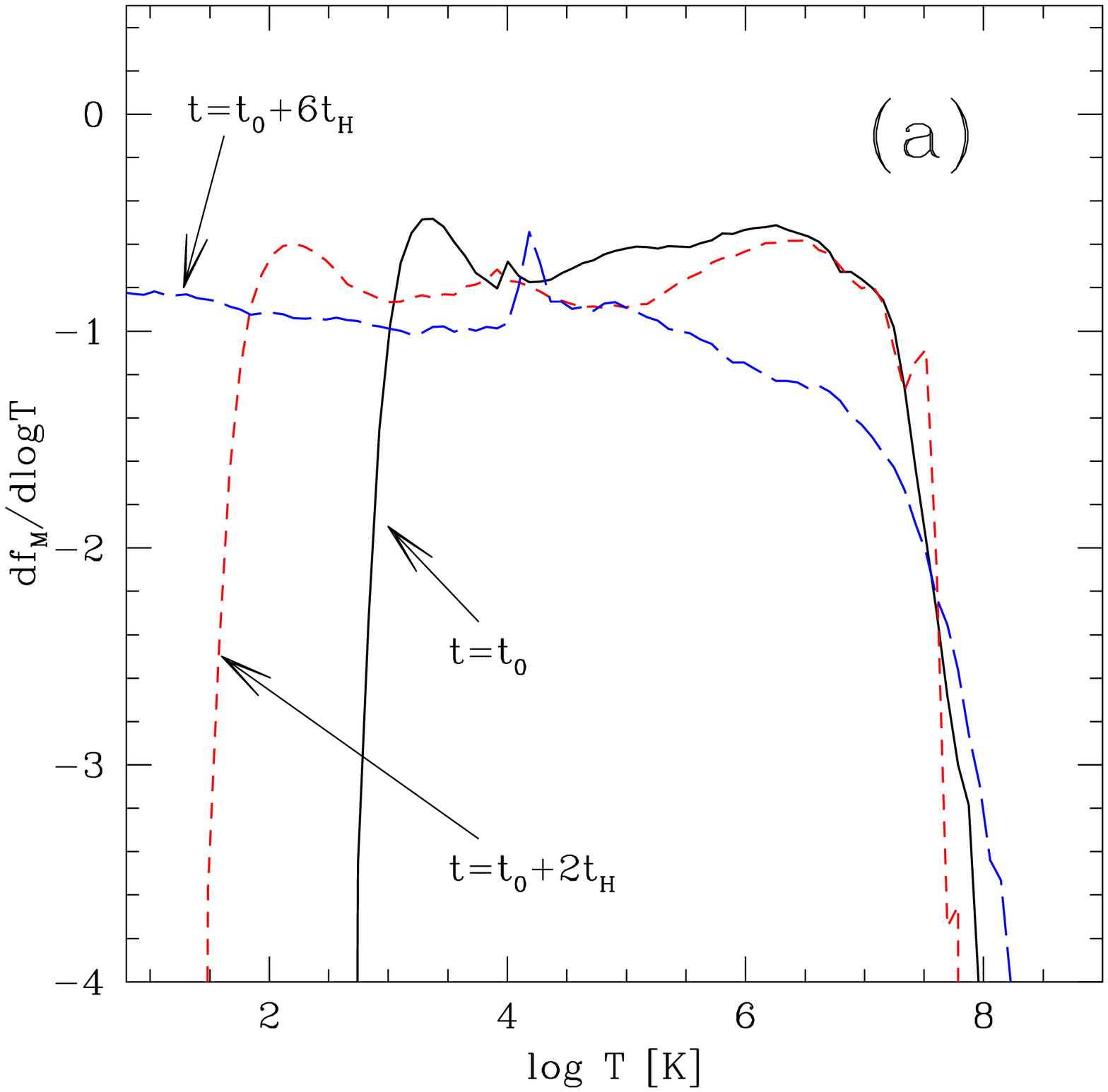}}\\
\hspace{0.3cm}
\resizebox{10cm}{!}{\includegraphics{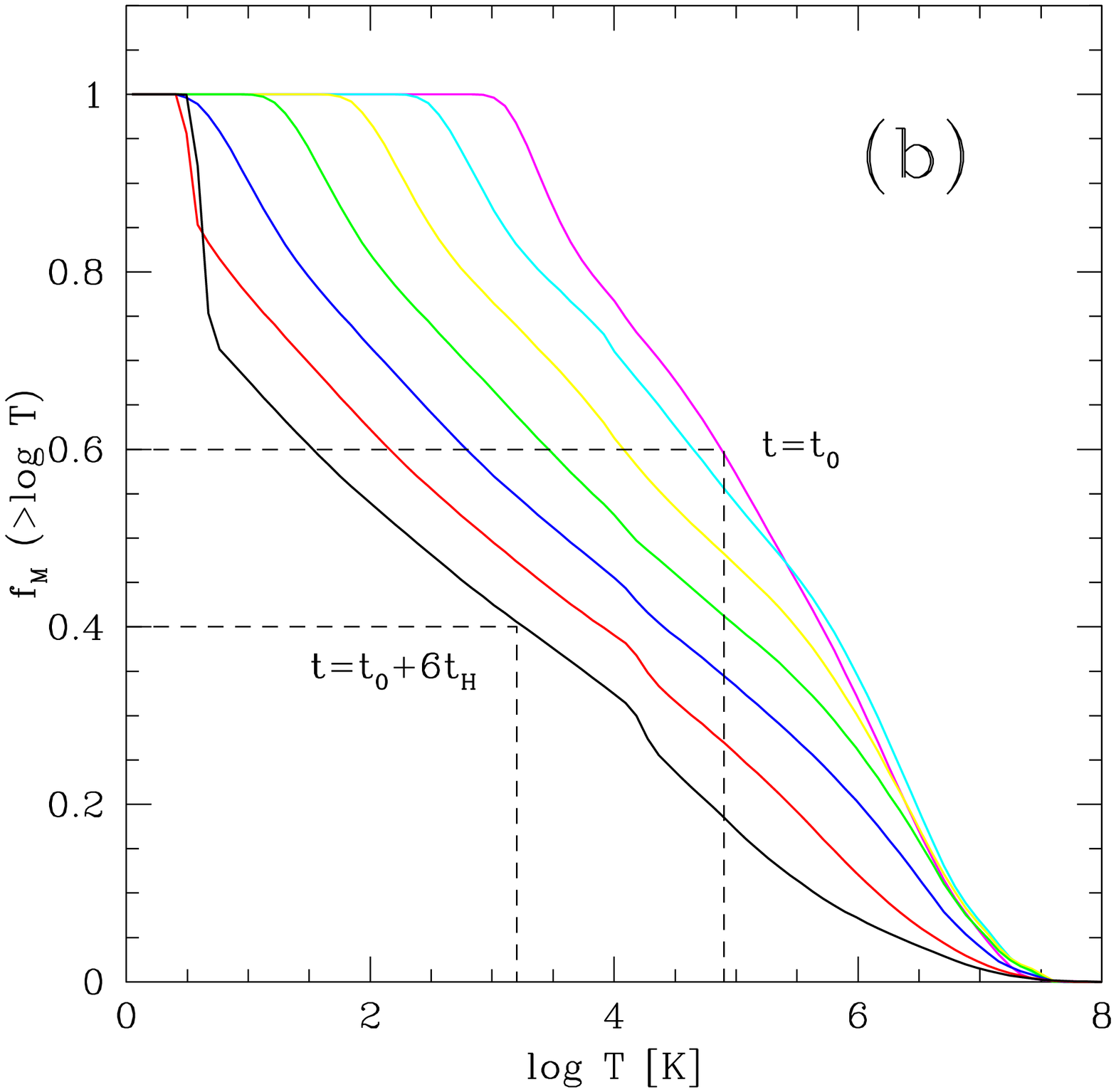}}\\%
\caption{{\it Panel (a):} Mass-weighted, differential distribution function
of gas temperature $T$ (in K), i.e. $df_M/d\log T$, at $t=t_0$, $t_0+2\tH$, 
and $t_0+6\tH$.  The distribution is truncated artificially at 
$\log (T/{\rm K})=0.7$.  {\it Panel (b):} Cumulative mass fraction of gas 
with temperature above a certain value for $t=t_0$, $t_0+2\tH$, $t_0+3\tH$, 
$t_0+4\tH$, $t_0+5\tH$, and $t_0+6\tH$ from right to left. See the text for 
the description of the dashed lines.
\label{temp.eps}}
\end{center}
\end{figure}

\Fig{temp.eps} shows the mass-weighted distribution function of gas as a 
function of the logarithm of the temperature $df_M/d\log T$  [panel (a)], 
and the cumulative mass fraction of gas $f_M(>\log T)$ with a temperature 
larger than a certain value [panel (b)].

Panel (a) shows that the minimum IGM temperature decreases as a function of
time due to the expansion of the Universe, while the maximum temperature
stays constant at $T=10^8$~K due to the lack of growth in the potential
well depth of dark matter halos.  The artificial cutoff at the temperature
floor of $T=5$~K is set by hand in the simulation.  

Panel (b) implies that the mass fraction of gas with $T>10^5$~K is roughly
60\% at $t=t_0$, and that this fraction decreases to 20\% at
$t=t_0+6\tH$. The mass fraction of gas with $T>10^{3.2}$~K is 40\% at
$t=t_0+6\tH$ (corresponding to the mass fraction above $1+\del=10^3$ in
\Fig{density.eps}).

\Fig{temp_evolve.eps} shows the evolution of the maximum temperature 
for gas with a fixed gas mass fraction of 40\% in lower density regions
as a function of the logarithm of the scale factor. This quantity measures 
the evolution of the temperature at $f_M(\log T)=0.6$ in panel (b) of 
\Fig{temp.eps} because \Fig{temp.eps} was a cumulative plot. 
\Fig{temp_evolve.eps} indicates that the temperature of the diffuse IGM 
cools down roughly according to the adiabatic expansion law, 
$T\propto a^{-2}$, as illustrated by the dashed line.
We confirmed that the result doesn't change very much even in the 
`max-UV field' run which we described in Section~\ref{section:simulation}.

\begin{figure}
\vspace{3cm}
\begin{center}
\epsfig{file=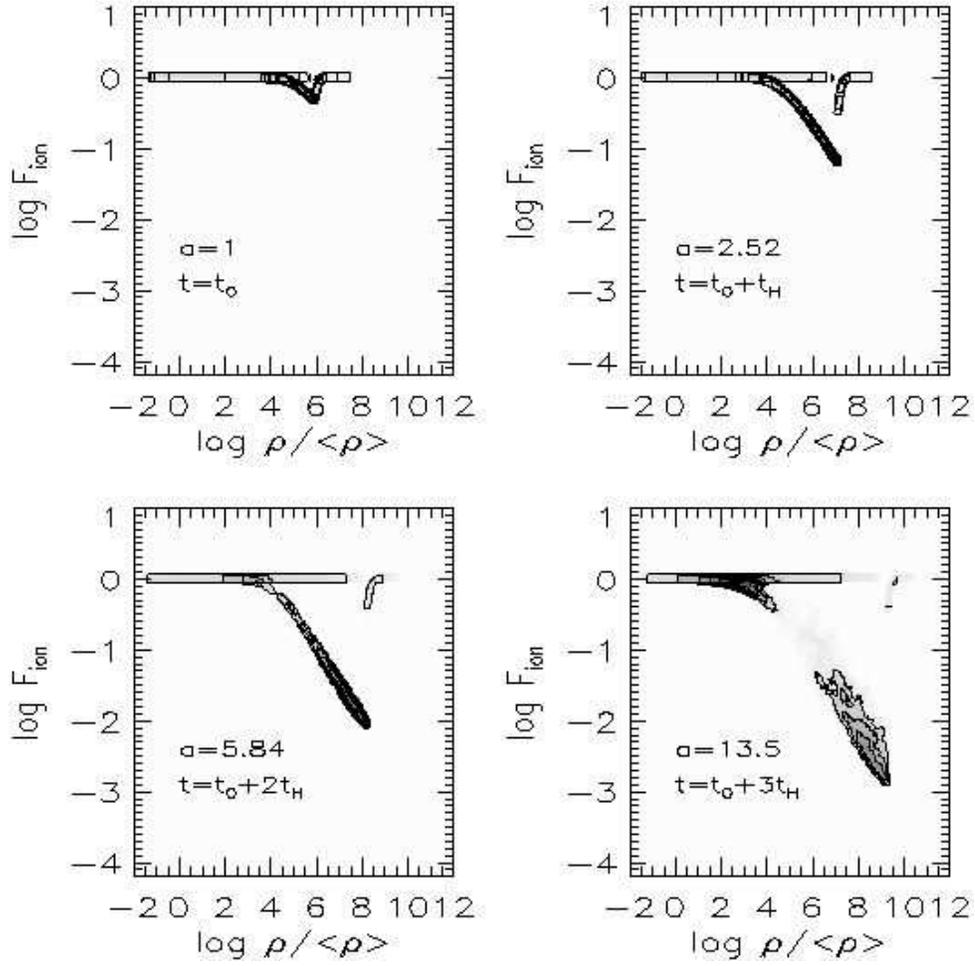,height=5in,width=5in, angle=0}
\caption{Mass-weighted mean temperature of the gas in the Universe, 
$\avg{T}$, as a function of the scale factor, $a=(1+z)^{-1}$. 
The upper horizontal axis indicates the corresponding redshift, $z$.} 
\label{mean_temp.eps}
\end{center}
\end{figure}

\Fig{mean_temp.eps} shows the mean (mass-weighted) temperature of all 
the gas in the Universe, $\avg{T}$, as a function of the scale factor, 
$a=(1+z)^{-1}$. The upper horizontal axis indicates the corresponding
redshift, $z$. It can be seen that the beginning (redshifts $z>20$) of the 
thermal history of the diffuse IGM are dominated by the adiabatic cooling. 
This is also true for the end of the thermal history of the diffuse IGM
as we already showed in \Fig{temp.eps}.
In between these two cooling phases, there is an epoch of structure formation 
during which the IGM is heated through large-scale shocks or the radiation 
emitted by stars and quasars.  The value of $\avg{T}$ at late times is 
dominated by the gas trapped in virialized dark matter halos which will cool 
only slowly with cosmic time. Overall, we find that $\avg{T}$ will peak at 
a value of $\sim 2\times 10^6$~K at $a\sim 2.5$, only one Hubble time 
($\sim 14$ billion years) from the present epoch. The appearance of the 
peak value of $\avg{T}$ in the relatively near future simply reflects the 
lack of subsequent growth in the mass function of virialized halos.


\section{Ionization Fraction}
\label{section:ionizefrac}

Finally, we consider the ionization fraction of the cosmic gas, which
we define as
\begin{eqnarray}
\Fion = \frac{n_e}{n_e + n_n},
\end{eqnarray}
where $n_e$ is the electron number density, and $n_n$ is the total number
density of neutral atoms. Since our simulation includes only the ionization
of hydrogen and helium, $n_n=n_{\rm HI}+n_{\rm HeI}$.  With this
definition, $\Fion \rightarrow 1$ for a fully ionized gas and $\Fion
\rightarrow 0$ for fully neutral gas.

\begin{figure}
\begin{center}
\epsfig{file=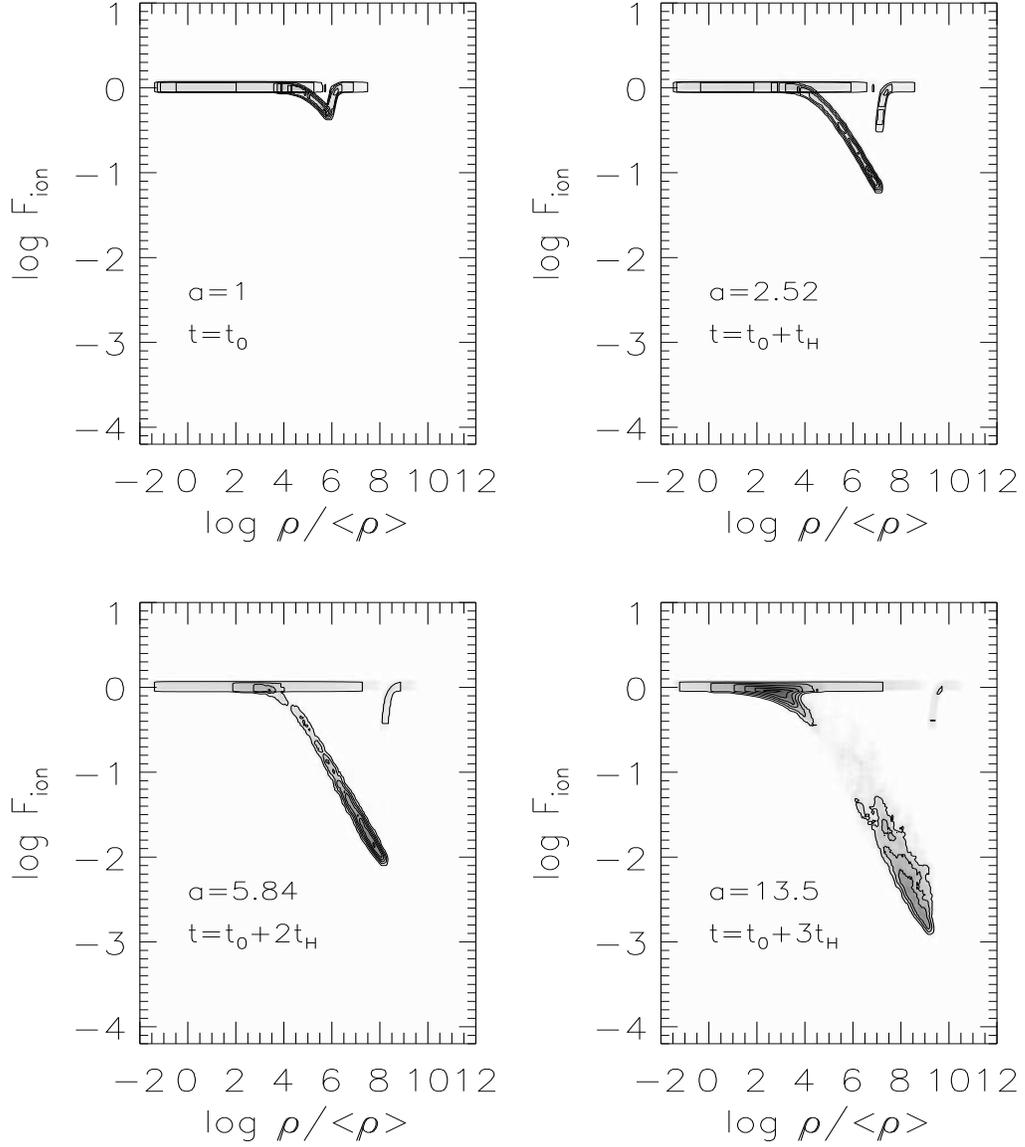,height=6in,width=5.3in, angle=0}
\caption{Mass-weighted distribution of ionization fraction of gas as a
function of gas overdensity (relative to the mean value $\avg{\rho}$) at
$t=t_0$, $t_0+\tH$, $t_0+2\tH$, and $t_0+3\tH$. The six contour levels 
are equally spaced in logarithmic intervals between the minimum and 
maximum values of the gas mass distribution on the plane.
\label{ionizefrac.eps}}
\end{center}
\end{figure}

\Fig{ionizefrac.eps} shows the mass-weighted distribution of the ionization
fraction as a function of the comoving gas overdensity.  The contours and
the gray levels are displayed on six equal logarithmic intervals in between 
minimum and maximum values of the projected gas mass density.  The top left 
panel indicates that most of the gas at $t=t_0$ is fully ionized
except for a pockets of neutral gas in star forming regions at $\log
\rho/\avg{\rho}\sim 6.0$. At even higher densities
($6<\log\rho/\avg{\rho}<8$), the gas is ionized again due to supernovae
feedback.

The neutral fraction in star forming regions (corresponding to the
narrow contour structure extending down to the lower-right corner of each
panel at $\log\rho/\avg{\rho}>6$) continues to grow until $t\approx
t_0+2\tH$, but diminishes afterwards as the neutral gas is consumed by star
formation. Additional infall onto virialized halos is suppressed in an
exponentially expanding universe. After $t=t_0+3\tH$, recombination is only
important in the dense cores of virialized halos, and the IGM remains
ionized as it continues to expand and cool. The distribution of $\Fion$ 
in the IGM does not evolve at later times.

\begin{figure}
\begin{center}
\epsfig{file=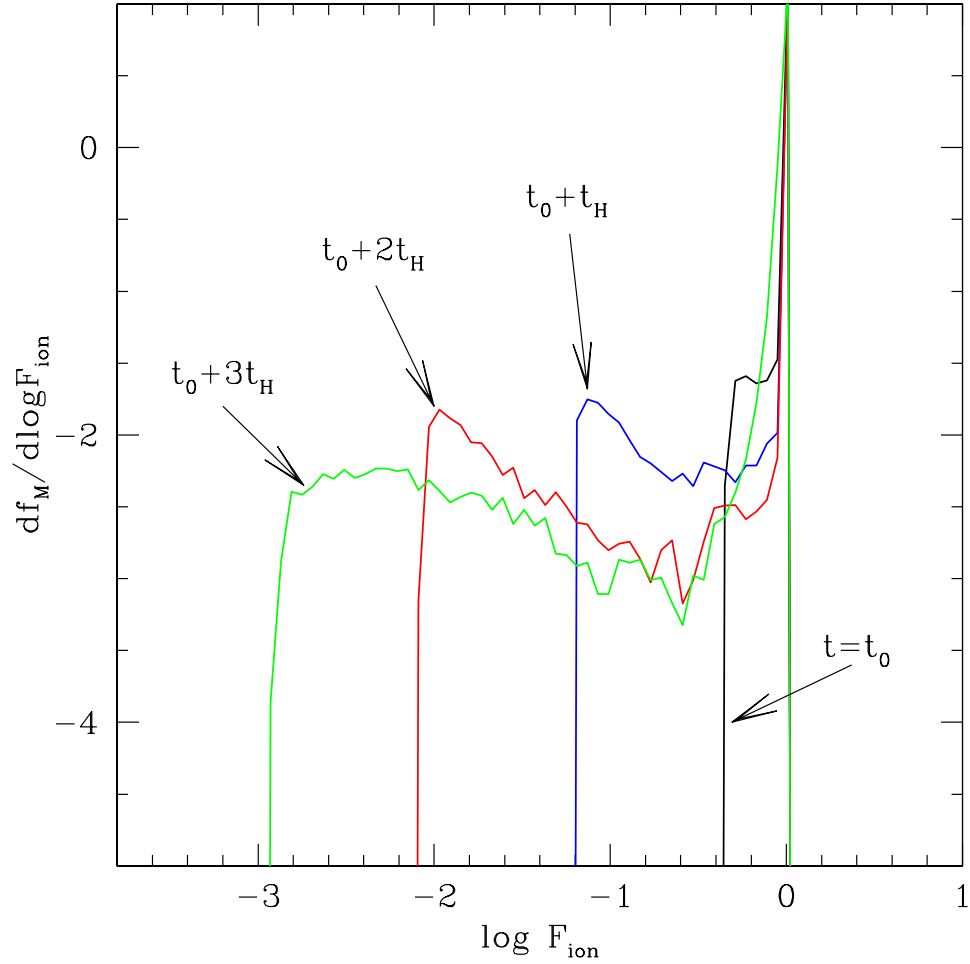,height=5in,width=5in, angle=0}
\caption{Mass-weighted differential distribution function
of the ionization fraction of the gas, i.e. $df_M/d\log \Fion$, 
at $t=t_0$, $t_0+\tH$, $t_0+2\tH$, and $t_0+3\tH$.
\label{dist_ionizefrac.eps}}
\end{center}
\end{figure}

\Fig{dist_ionizefrac.eps} shows the mass-weighted distribution function of
the gas mass as a function of its ionization fraction, i.e. $df_M / 
d\log \Fion$.  The lower end of the distribution gradually extends to 
lower $\Fion$ values between $t=t_0$ and $t=t_0+3\tH$, although most of the 
mass is still fully ionized ($\log \Fion = 0.$). At $t=t_0+3\tH$, 99\% of 
the gas mass has an ionization fraction higher than 80\%.  Subsequently, 
the $\Fion$ distribution freezes outside the dense cores of virialized halos.  

In our simulation, we have assumed that the gas is optically thin and in
ionization equilibrium when computing the abundance of different ionic
species. This assumption is of course only valid in the high density region
where the collisional ionization and recombination time scales are shorter
than the Hubble time. We find that the ionization fraction evolve only for
$\log(\rho/\avg{\rho}) > 4$ until $t=t_0+3\tH$. In this regime of
densities, the recombination rate is still higher than $1/t$, and the
assumption of ionization equilibrium is justified.  After $t=t_0+3\tH$, the
recombination time-scale becomes much longer than the Hubble time for most
of the plotted range of densities, and the distribution of ionization
fraction of the IGM is expected to freeze for gas that has not assembled
into dark matter halos.


\section{Conclusions}
\label{section:conclusion}

We have simulated the future evolution of the intergalactic medium in a
universe dominated by a cosmological constant, focusing on the
overdensity, temperature, and ionization fraction of the cosmic gas.

\begin{figure}
\vspace{3cm}
\begin{center}
\epsfig{file=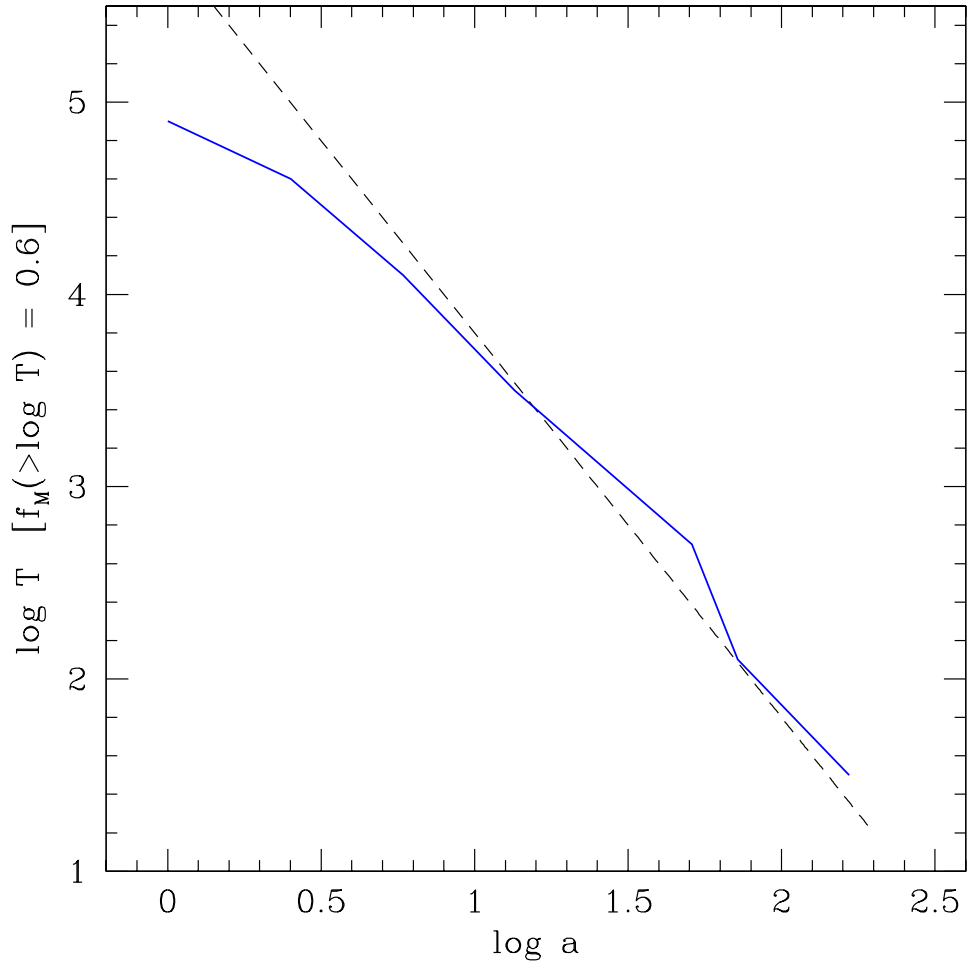,height=5in,width=5in, angle=0}
\caption{Evolution of the maximum temperature for gas with a fixed mass
fraction of 40\% in lower density regions as a function of the logarithm of
the scale factor. This quantity measures the evolution of the temperature 
at the fixed value of $f_M(>\log T)=0.6$ in panel (b) of \Fig{temp.eps} 
because \Fig{temp.eps} is a cumulative plot.  The curve indicates that 
the temperature of the diffuse IGM cools roughly according to the adiabatic
 expansion law $T\propto a^{-2}$, shown by the dashed line.}
\label{temp_evolve.eps}
\end{center}
\end{figure}

We have found that within a few Hubble times from the present epoch, the
baryons will split into two major phases: one phase of low-density,
low-temperature, diffuse IGM which cools adiabatically (with $T\propto 
a^{-2}$; see \Fig{temp_evolve.eps}), and a second phase of high-density,
high-temperature gas in virialized dark matter halos which cools more
slowly by up to two orders of magnitude [see Equation (\ref{eq:cooling})]. 
The mass fraction of gas which is confined in virialized dark matter halos 
(defined as regions with an overdensity larger than $\sim 200$) converges 
to $\sim 40\%$ at late times, about twice its calculated value at the 
present epoch.

The simulated maps of gas temperature show that the large-scale filaments
disperse and merge with the low-temperature IGM background after a few
Hubble times, and only the `island universes' of virialized gas maintain
their physical structure. Although these islands are linked by filaments of
dark matter and gas in comoving coordinates, the filaments rarefy and
disperse in physical coordinates due to the exponential temporal growth of
the cosmic scale factor, $a$.  We do not find evidence for fragmentation of
the baryons in the IGM at later times above the mass-scale of $M_{gas} \sim
5\times 10^{11}\himsun$.

The recombination time of the expanding IGM exceeds the Hubble time by a
factor that grows rapidly in the future, and so most of the IGM gas remains
ionized. After three Hubble times from the present epoch, 99\% of the gas
mass maintains an ionization fraction above 80\%.  If the Universe is
indeed dominated by a true cosmological constant, then the diffuse IGM
outside virialized dark matter halos will get extremely cold but remain
highly ionized.


\ack

We thank Volker Springel and Lars Hernquist for helpful discussions 
and for allowing us to use the
updated parallel version of the GADGET code which is described in
\citet{SH02, SH03a, SH03b}.  This work was supported in part by the grants
ATP02-0004-0093 from NASA and AST-0071019, AST-0204514 from NSF for AL.
The simulations were performed at the Center for Parallel Astrophysical
Computing at Harvard-Smithsonian Center for Astrophysics.



\end{document}